\documentclass[superscriptaddress,twocolumn]{revtex4-1}
\usepackage{amsmath}
\usepackage{amssymb}
\usepackage{graphicx}
\usepackage{hyperref}
\usepackage{amsfonts}
\usepackage{bbm}
\usepackage{datetime}
\usepackage[english]{babel}

\usepackage[utf8]{inputenc}

\usepackage{color}
\usepackage{xcolor}
\usepackage{esint}
\usepackage{bm}
\usepackage{hyperref}
\longdate

\newcommand{\quotes}[1]{``#1''}

\makeatletter
\newcommand*{\rom}[1]{\expandafter\@slowromancap\romannumeral #1@}
\makeatother

\begin{document}
\title{Spectral energy analysis of bulk three-dimensional active nematic turbulence} 
\author{Žiga Krajnik}
\affiliation{Faculty of Mathematics and Physics, University of Ljubljana, Jadranska 19, 1000 Ljubljana, Slovenia.}
\author{\v{Z}iga Kos}
\affiliation{Faculty of Mathematics and Physics, University of Ljubljana, Jadranska 19, 1000 Ljubljana, Slovenia.}
\affiliation{Present address: Department of Mathematics, Massachusetts Institute of Technology, Cambridge, MA 02139-4307, USA.}
\author{Miha Ravnik}
\email{miha.ravnik@fmf.uni-lj.si}
\affiliation{Faculty of Mathematics and Physics, University of Ljubljana, Jadranska 19, 1000 Ljubljana, Slovenia.}
\affiliation{Jo\v{z}ef Stefan Institute, Jamova 39, 1000 Ljubljana, Slovenia.}

\begin{abstract}
We perform energy spectrum analysis of the active turbulence in 3D bulk active nematic using continuum numerical modelling. Specifically, we calculate the spectra of two main energy contributions---kinetic energy and nematic elastic energy---and combine this with the geometrical analysis of the nematic order and flow fields, based on direct defect tracking and calculation of autocorrelations. We show that the active nematic elastic energy is concentrated at scales corresponding to the effective defect-to-defect separation, scaling with activity as $\sim\zeta^{0.5}$, whereas the kinetic energy is largest at somewhat larger scales of  typically several 100 nematic correlation lengths. Nematic biaxiallity is shown to have no role in active turbulence at most of length scales, but  can affect the nematic elastic energy by an order of magnitude at scales of active defect core size. The effect of an external aligning field on the 3D active turbulence is explored, showing a transition from an effective active turbulent to an aligned regime. The work is aimed to provide a contribution towards understanding active turbulence in general three-dimensions, from the perspective of main energy-relevant mechanisms at different length scales of the system.  
\end{abstract}
\date{\today}
\maketitle

\section{Introduction}

Active turbulence is the major result of microscopic activity and nematic orientational ordering in diverse active nematic fluids at low Reynolds numbers, ranging from microfilaments and microtubules, bacterial suspensions to epithelia~\cite{WuK_Science355_2017,WensinkHH_ProcNatlAcadSci109_2012,BlanchMercaderC_PhysRevLett120_2018}. It is characterised by the emergence of active topological defects, i.e. regions of lost microscopic order that are steadily generated, annihilated and transformed in time in a complex and irregular manner. Topological defects---in three spatial dimensions: lines, points, loops, or walls---are regions of broken orientational order characterized by a locally discontinuous director (i.e. the direction of the average alignment) and a reduced degree of order~\cite{AlexanderGP_RevModPhys84_2012}. In active materials, the formation and spatio-temporal coherence of topological defects depend strongly on the active material being either extensile or contractile~\cite{ElgetiJ_SoftMatter7_2011} and on the concentration of the active agents~\cite{SokolovA_PhysRevLett98_2007}. 

Despite the diversity of these systems, there is strongly recurring behaviour of pattern formation~\cite{DombrowskiC_PhysRevLett93_2004,GiomiL_PhysRevLett106_2011,CarballidoLandeiraJ_Langmuir31_2015,SokolovA_PhysRevX9_2019,KruseK_PhysRevLett92_2004} and the appearance of topological defects~\cite{SankararamanS_PhysRevLett102_2009,ThampiSP_PhilTransRSocA372_2014,VicsekT_PhysRevLett75_1995,TonerJ_PhysRevLett75_1995,CoparS_PhysRevX9_2019}. Indeed, especially at high activities and densities, the response of active materials becomes generically similar; the active agents start to collectively move in irregular and chaotic time changing patterns, which is known as active turbulence~\cite{WuK_Science355_2017,ThampiSP_PhilTransRSocA372_2014,SanchezT_Nature491_2012,WensinkHH_ProcNatlAcadSci109_2012}. Experimentally, the key challenge is to observe the turbulent time-changing patterns (e.g. of microtubules or bacteria), fast enough and in principle in 3D, as it requires fast dynamic tracking of active agent orientations as well as their motility~\cite{dogicArxiv}. Theoretically, the challenge is in establishing a clear way for analysing such 3D time-changing active nematic fields, as simply plotting velocity or orientational fields of active nematics (typical outputs of the active nematic theory) reveals limited fundamental information~\cite{GiomiL_PhysRevX5_2015}.

The dynamics of active systems~\cite{MarchettiMC_RevModPhys85_2013,DoostmohammadiA_NatCommun9_2018} is closely related to liquid crystal hydrodynamics~\cite{book-deGennes,VoituriezR_EurophysLett70_2005,PengC_Science354_2016,MushenheimPC_SoftMatter10_2013,sohn2019optically,C8SM00612A}. Recently, studies of active nematic layers show the importance of geometry and topology in active nematics, as also revealed via active turbulence~\cite{SanchezT_Nature491_2012,GiomiL_PhysRevX5_2015,ThampiSP_PhilTransRSocA372_2014}. Spectral densities and correlation functions are used to characterise the systems~\cite{GiomiL_PhysRevX5_2015}, with current knowledge on active turbulence  primarily based on analysis of 2D and quasi-2D layers  of active nematics, with clear interest in full 3D spatial configurations. The chaotic motion of turbulence necessitates a probabilistic description in which a convenient tool, amenable to analysis, is the spectrum of an apparently random process~\cite{book-Frisch}. 
The motivation of this work is to study active turbulence---in full 3D---by analysing the system's energy spectra, drawing from the analogies with, though profoundly different, classical high Reynolds number turbulence~\cite{UrzayJ_JFluidMech822_2017}, where energy spectrum analysis (such as the renowned Kolmogorov theory~\cite{KolmogorovAN_ProcRSocA434_1991}) proves to be the strongest methodological approach to characterize turbulence. In active turbulence, the system is at low Reynolds numbers with little effects of inertia (which differently, are central in high-Re turbulence) but the analogy is in having to consider irregularly (chaotic) changing material fields, distinctly, the material flow and the orientational order.

In this Article, we study 3D bulk chaotic low Reynolds number dynamic state in an active nematic liquid crystal frequently known  as active turbulence, using continuum modelling. We quantify the two main energy mechanisms that determine the active turbulence---the kinetic energy and the nematic elastic energy---across different length scales by  computing the corresponding energy spectra. We show that the maximum in the elastic spectrum scales with the effective concentration of defect lines, whereas at small physical scales, the elastic spectrum is  up to multiplication by a constant independent of activity  and correspond to the molten defect lines. The active nematic elastic energy (of nematic orientational ordering) is mainly contained on scales comparable to the average distance between defect lines, typically of 10--100 nematic correlations lenghts ($\xi$) dependent on activity, as further supported by geometrical analysis. Somewhat differently, the kinetic energy (of the flow field) is largest at  scales relatively larger than elastic energy, i.e. of  typically several 100 nematic correlation lengths. The role of nematic biaxiallity in 3D active nematic turbulence is also considered showing that it can affect the nematic elastic energy by factor of 10, at length scales comparable to defect core size (i.e. below $10\,\xi$). We also demonstrate the effect of an aligning homogeneous external (electric) field on nematic turbulence and observe a transition to a frozen completely-aligned phase at an activity dependent critical field strength. More generally, the  energy spectra analysis of different active materials in the active turbulence regime could  be possibly used for  associating mutually different active matter systems into common groups, and also serve as the basis for linking experiments with appropriate theoretical (computational) models. Note also that raw data of selected 3D bulk active nematic turbulence simulations presented in this work (time evolution of nematic tensor $Q_{ij}$ and the velocity field $u_i$) is made available in Ref.~\cite{ziga_kos_2019_3541954}.

\section{Theory and modelling}
Bulk three-dimensional active nematic turbulence is approached by a combination of numerical modelling based on generalised Beris-Edwards equations for active nematics~\cite{book-Beris, HatwalneY_PhysRevLett92_2004} and material analysis, especially using spectral (Fourier) analysis of the main material fields, as further explained below. 

\subsection{Mesoscopic modelling of active nematics}
Mesoscopic modelling of active nematic is based on coupled 3D spatially and in-time varying fields for fluid velocity $\vec{u}$, density $\rho$, and the nematic tensor order parameter $\mathbf{Q}$ with the largest eigenvector---the director $\vec{n}$---characterizing the main local ordering axis and the corresponding eigenvalue characterizing the degree of order $S$. This is a rather established approach and was shown in the  literature to be able to explain key features in the experiments of active nematics~\cite{ThampiSP_PhilTransRSocA372_2014, GiomiL_PhilTransRSocA372_2014,jackArxiv}. The model builds on governing equations for the time evolution equation for nematic order, generalized Navier-Stokes equation and the incompressibility condition: 
\begin{eqnarray}
(\partial_t+u_k\partial_k)Q_{ij}-S_{ij}&=&\Gamma H_{ij} ,\label{eqqten} \\
\rho(\partial_t+u_k\partial_k)u_i&=&\partial_j \Pi_{ij}, \label{eqflow}\\
\partial_ku_k&=&0,\label{u_eq}
\end{eqnarray}
where $\partial_t$ is the partial derivative with respect to time, $\partial_i$ is partial derivative with respect to a Cartesian spatial coordinate, and $\Gamma$ is the rotational diffusion coefficient. Summation over repeated indices is assumed. The non-equilibrium dynamics of active nematics is driven by the active stress, which enters into the equation for the total stress tensor $\Pi_{ij} = \Pi_{ij}^\text{active}  + \Pi_{ij}^\text{passive}$:
\begin{eqnarray}
\Pi_{ij}^\text{active}&=&-\zeta Q_{ij},\\\label{eqActStress}
\Pi_{ij}^\text{passive}&=&2\eta D_{ij} + 2\lambda(Q_{ij}+\delta_{ij}/3)(Q_{kl}H_{lk})\nonumber\\
&-&\lambda H_{ik}(Q_{kj}+\delta_{kj}/3) -\lambda (Q_{ik}+\delta_{ik}/3)H_{kj}\nonumber\\
&+&Q_{ik}H_{kj}-H_{ik}Q_{kj}-\partial_i Q_{kl}\frac{\delta \mathcal{F}}{\delta\partial_j Q_{lk}}-p\delta_{ij},
\label{active_stress}
\end{eqnarray}
where $\zeta$ is the activity parameter, $\eta$ an isotropic contribution to the viscosity, $D_{ij}=(\partial_iu_j+\partial_ju_i)/2$, $\lambda$ is the alignment parameter, $p$ is pressure, and $H_{ij}=-\frac{\delta\mathcal{F}}{\delta Q_{ij}}+(\delta_{ij}/3)\mathrm{Tr}\frac{\delta\mathcal{F}}{\delta Q_{kl}}$ is the molecular field calculated from the free energy $\mathcal{F}$ with density of
\begin{eqnarray}
f&=&\frac{L}{2}(\partial_k Q_{ij})^2-\frac{1}{3}\epsilon_0\epsilon_\text{a}^\text{mol}E_iE_j \label{free_en}\\
&+&\frac{A}{2}Q_{ij}Q_{ji}+\frac{B}{3}Q_{ij}Q_{jk}Q_{ki}+\frac{C}{4}(Q_{ij}Q_{jk})^2\nonumber
\end{eqnarray} 
with the elastic constant $L$, vacuum permittivity $\epsilon_0$, relative molecular anisotropic permittivity $\epsilon_\text{a}^\text{mol}$, electric field $\vec{E}$, and material constants  $A$, $B$, and $C$.
The advection term $S_{ij}$ is defined as:
\begin{eqnarray}
&&S_{ij}=-2\lambda(Q_{ij}+\delta_{ij}/3)(Q_{kl} \partial_{k}u_l)\\
&&+(\lambda D_{ik}-\Omega_{ik})(Q_{kj}+\delta_{kj}/3)\nonumber \\
&&+(Q_{ik}+\delta_{ik}/3)(\lambda D_{kj}+\Omega_{kj}),\nonumber
\end{eqnarray}
where $\Omega_{ij}=(\partial_i u_j-\partial_j u_i)/2$. 

Equations of active nematodynamics (\ref{eqqten}--\ref{u_eq}) include two intrinsic length scales that govern the behaviour of the active turbulence. These are the nematic correlation length $\xi$---also present in passive nematics---and active lengthscale $\xi_\zeta$. The nematic correlation length $\xi=\sqrt{\frac{L}{A+BS_\text{eq}+9CS_\text{eq}^2/2}}$ ($S_{eq}$ is the equilibrium nematic degree of order) is given as an effective ratio between nematic elasticity and material ordering (in view of the nematic degree of order) and measures the effective thickness of nematic topological defects. Differently, active length $\xi_\zeta=\sqrt{L/|\zeta|}$  measures the effective ratio between nematic elasticity and the strength of the activity (i.e. active stress).  We will be considering a 3D bulk active nematic; but naturally, any geometrically confinement or presence of external fields  can also introduce additional length scales relevant in the system. Time evolution of active nematic can be measured in units of nematic intrinsic timescale $\tau_N=\xi_{N}^{2}/ \Gamma L$.

The coupled set of Equations (\ref{eqqten}-\ref{u_eq}) is solved by a hybrid lattice Boltzmann method, consisting of a finite difference approach for the Q-tensor time evolution (Eq.~\ref{eqqten}) and the 19 velocity lattice Boltzmann method with Bhatnagar-Gross-Krook collision operator for the incompressibility and Navier-Stokes equations. The simulations are performed in a rectangular simulation box  with a cubic lattice on 405 x 405 x 405 mesh points. Periodic boundary conditions for both $Q_{ij}$ and $u_i$ are used in all three spatial directions. The mesh resolution is $\Delta x=1.5\xi$. The following values for the material parameters were used in the simulation: $A=-0.190\,L/\xi^2$, $B=-2.34\,L/\xi^2$, $C=1.91\,L/\xi^2$, $\rho=0.0275/L\Gamma^2$, and $\eta=1.38/\Gamma$. We take the flow aligning parameter $\lambda = 1$ which  corresponds to the dynamics of rod-like active nematic building blocks and in the limit of passive nematic fluids would correspond to flow-aligning regime. The value of $\lambda$ (i.e. aligning vs tumbling) was shown to be important when active nematics interact with surfaces \cite{ThijssenK_SoftMatter} and in the transition into a turbulent flow \cite{Chandragiri_SoftMatter}.  Within the simulation regimes, Reynolds number was taken to be $\text{Re}\sim 1$, as low as reasonably achievable in such type of rather large mesoscopic volume simulations, inline with comparable studies in the literature \cite{HemingwayEJ_SoftMatter12_2016}. The starting state for the simulations was a zero initial velocity field. The initial $Q$ field  was obtained using the uniaxial approximation $Q_{ij} = \frac{3S_\text{eq}}{2}(n_i n_j - \frac{1}{3}\delta_{ij})$ where $n_i$ was a unit vector with a uniform random distribution in space and $S_\text{eq}=0.533$. The simulations were run for $10^5$ time steps (each time step corresponds to $0.0225\,\tau_N$): first 60000 time steps ($1350\,\tau_N$) were used to reach the dynamic steady state (of the active turbulence) and the remaining 40000 steps ($900\,\tau_N$) were used in the calculation of time-averaged quantities.

\begin{figure*}
\centering \includegraphics[width=\textwidth]{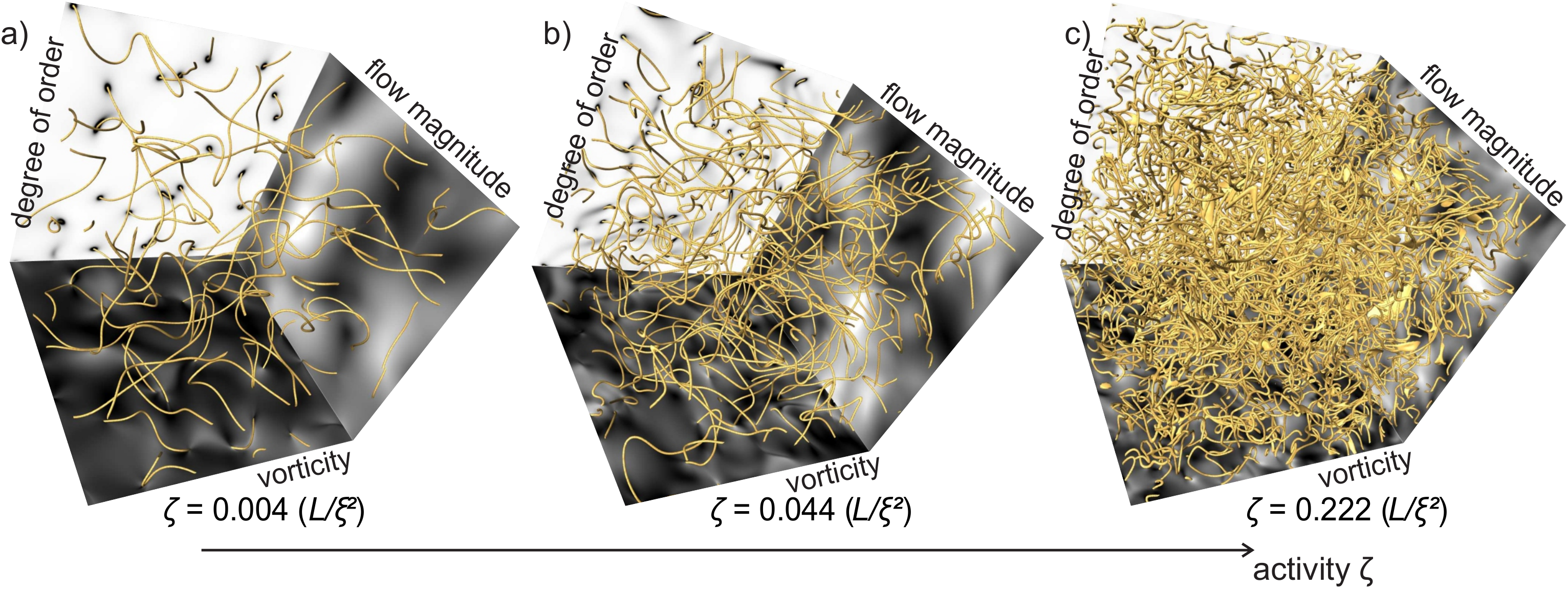}
\caption{Active turbulence in bulk 3D active nematic at three different activities $\zeta$. Defect lines are drawn (in yellow) as isosurfaces of the nematic degree of order at $S= 0.4$. Grayscale colourmaps show profiles of the degree of order $S$, flow magnitude, and vorticity in the given cross-section planes.}
\label{Fig1}
\end{figure*}

\subsection{Spectral analysis}

The ordering and flow profiles of 3D active turbulence patterns can be analysed by using spectral analysis, which gives insight into the material structure over different length scales. Specifically, we analyse the spectra of main energy contributions in the system---kinetic energy and nematic elastic energy---spectra, motivated by the analogy to the kinetic spectra that are used to study inertial (high Reynolds number) turbulence. We should comment that the spectra are introduced as  one-dimensional $E(k)$ (although $k$ is the length of the full 3D vector $\bm{k}$), as the active turbulence is isotropic; therefore, the one dimensional spectrum is sufficient as the signal in all directions in $k$ space is equivalent. In non-isotropic systems, in principle, 
full three dimensional spectra $E(\bm{k})$ have to be considered, if to extract all the information, whereas 1D spectra only provide an effective (averaged) view on the anisotropy.

The kinetic energy spectrum $E_{kin}$ is implicitly defined as: 
\begin{equation}
w_{kin} = \int_{0}^{\infty}E_{kin}(k) dk = \frac{1}{2} \frac{\rho}{V} \int  u_i(\mathbf{r})u_i(\mathbf{r}) d^D r, \label{def}
\end{equation}
where $w_{kin}$ is the average kinetic energy density, $\rho$ is density, $V$ is volume, $u_i(\mathbf{r})$ is velocity in direct space, $k$ is the magnitude of the wave vector in reciprocal space and $D$ is the space dimension. 
Alternatively, the kinetic energy spectrum can be written in the reciprocal (Fourier) space as: 
\begin{equation}
E_{kin}(k) = \frac{1}{2}\frac{\rho}{V} \oiint_{|\bm{k'}| = k} u_i(\bm{k'})u_i^*(\bm{k'}) \frac{dS_{\bm{k'}}}{(2\pi)^D},
\label{kin1}
\end{equation}
where the integration runs over a sphere in reciprocal space with a fixed radius $k$.
As an alternative to kinetic energy, enstrophy $\epsilon$ is another measure  that is often used to characterise velocity fields. It is defined as:
\begin{equation}
\epsilon = \int (\varepsilon_{ijk} \partial_j u_k)^2 d^Dr =  \int (\partial_i u_j)^2 d^Dr - \int (\partial_i u_i)^2 d^Dr  = \int (\partial_i u_j)^2 d^Dr,
\end{equation}
where $\varepsilon_{ijk}$ is the Levi-Civita tensor. In reciprocal space, the enstrophy spectrum can be written as:
\begin{equation}
E_{ens}(k) = k^2 \oiint_{|\bm{k'}| = k} u_i(\bm{k'})u_i^*(\bm{k'}) \frac{dS_{\bm{k'}}}{(2\pi)^D}.
\end{equation}
Note that in reciprocal space, the enstrophy is up to a pre-factor, the kinetic spectrum multiplied by $k^2$.

Analogously to the kinetic spectrum, we implicitly define the active nematic elastic spectrum---attributed to effective active nematic elasticity as:
\begin{equation}
w_{ela} = \int_{k = 0}^{\infty}E_{ela}(k) dk = \frac{1}{2} \frac{L}{V} \int (\partial_l Q_{ij} (\mathbf{r}))^2d^D r, \label{def2}
\end{equation}
where $w_{ela}$ is the nematic elastic energy per unit volume of the sample $V$, with the elastic spectrum in the reciprocal form expressed as: 
\begin{equation}
E_{ela}(k) = \frac{1}{2} \frac{L}{V} k^2 \oiint_{|\bm{k'}|=k}  Q_{mn}(\bm{k'}) Q_{mn}^*(\bm{k'})  \frac{dS_{\bm{k'}}}{(2\pi)^D},
\label{ela1}
\end{equation}
where the surface integral runs over a sphere in reciprocal space with fixed radius $k$.\\

To numerically evaluate the spectra we first compute the 3D discrete Fourier transform of $u_i$ or $Q_{ij}$ at a given time step $t$ and calculate the full three dimensional energy spectra $E^t(\bm{k})$ and then reduce it to a one dimensional spectrum by averaging the 3D spectrum over a thin spherical shell in reciprocal space and multiply the result with the volume of the shell. The discrete one dimensional spectrum at time $t$ is thus calculated as:
\begin{equation}
E^t(k_i) = \sum_{ ||\bm{k_j}| - k_i| < \Delta k}^{k_{max}} \frac{1}{N_i} {E^t(\bm{k}_j)} 4\pi k_i^2, \quad N_i =  \sum_{ ||\bm{k_j}| - k_i| < \Delta k} 1,
\label{discrete}
\end{equation}
where $k_i$ is the radius of the spherical shell in the reciprocal space, $\bm{k}_j$ are the 3D discrete wave vectors in reciprocal space and $\Delta k = \frac{\pi}{2 M}$ is the half-thickness of the shell ($M$ is the the length of the simulation box). The sum is cut off at $k = k_{max}=\pi/\Delta x$,
which corresponds to the Nyquist frequency and ensures that the whole spherical shell is contained in the reciprocal space. 
The spectra for a given set of parameters are then time averaged to obtain the final energy spectra (presented in the Results section):
\begin{equation}
E(k_i) = \frac{1}{N}\sum_{t=1}^N {E^t(k_i)}
\end{equation}
Note that, among other tests, we verified the performance of the above described method also by using the (high Reynolds number) Johns Hopkins Turbulence Database's \textit{forced isotropic turbulence} dataset~\cite{LiY_JournalofTurbulence9_2008, Perlman:2007:DET:1362622.1362654} and accompanying spectrum.\\

\section{Results}

\begin{figure}
\centering \includegraphics[width=8.7cm]{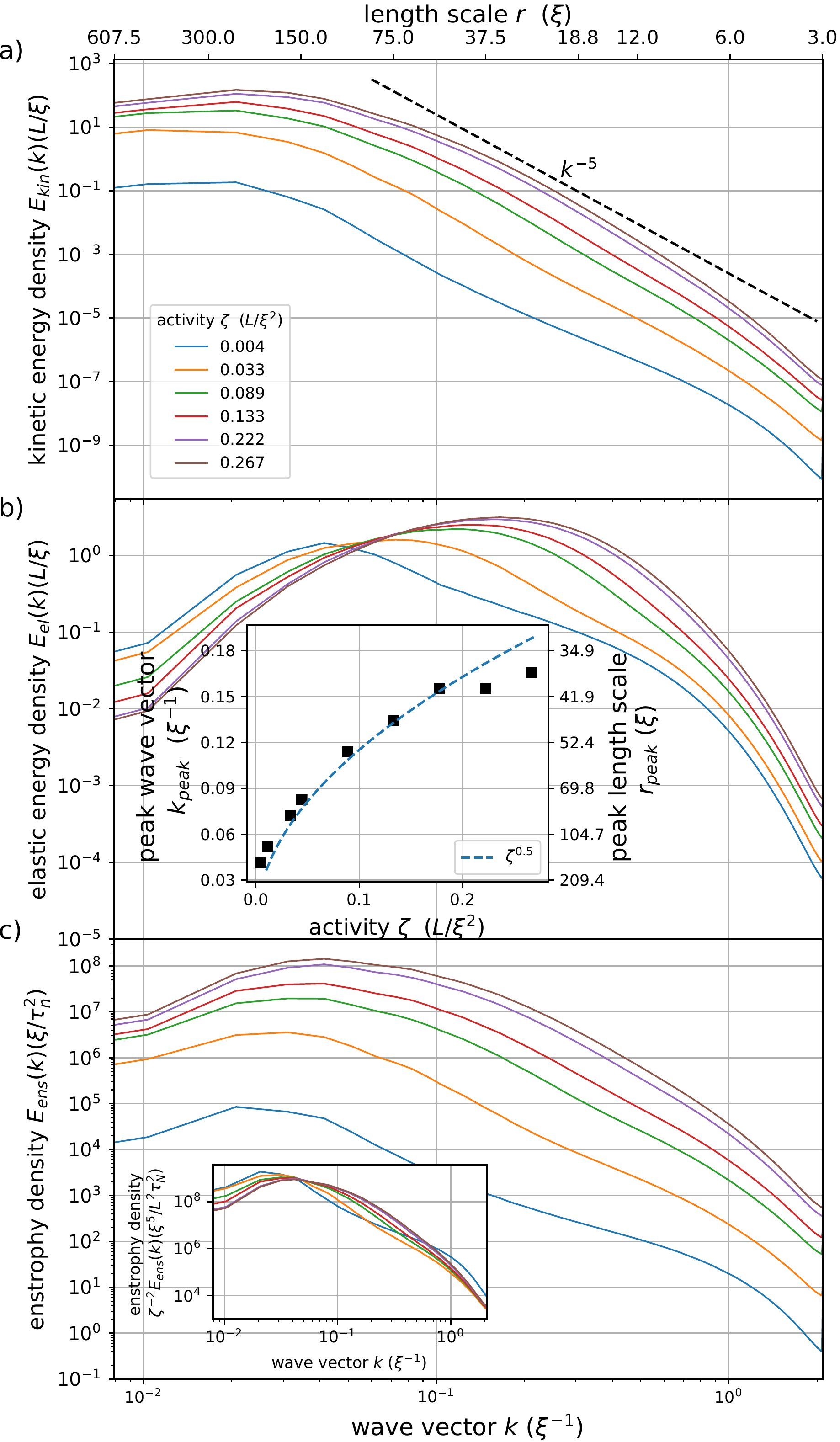}
\caption{Time averaged (a) kinetic, (b) elastic and (c) enstrophy spectra of bulk three dimensional active turbulence for different activities. Top axis gives the length $r$ in direct space in units of nematic correlation length $\xi$; bottom axis gives the length in reciprocal (Fourier) space of wave vectors. In kinetic energy spectrum,  dotted black line shows a $k^{-5}$ dependence as a guide to the eye. Elastic energy spectrum has a clear maximum whose position shifts with increasing activity as shown in the inset (b). Dotted blue line in the inset of elastic energy density spectra shows $\zeta^{0.5}$ curve (as guide to the eye) and simulation data is presented by black squares. Enstrophy spectra roughly scale with with $\zeta^{2}$, as shown in inset of (c).
} \label{Fig2}
\end{figure}

The dynamics state of the bulk 3D active nematic turbulence at different activities is shown in Fig.~\ref{Fig1}. The structure is characterised by a three-dimensional network of nematic defect lines that undergo constant dynamic re-morphing, as a results of constant energy input into the system via the active stress. The structure is irregular with no apparent order neither in the defect network nor in the velocity field and is conditioned by the mutual coupling between the active nematic orientation and the material flow. The defect network is constantly transforming in time, with segments of defect lines reconnecting, merging together and splitting apart, and defect loops undergoing annihilation. In a steady state, the defect density is generally constant and increases with higher activity. Animations of disclination dynamics are available in Supplementary Information (SI Video 1) and raw data of the presented simulations (time evolution of nematic tensor $Q_{ij}$ and the velocity field $u_i$) is made available in Ref.~\cite{ziga_kos_2019_3541954}.

\begin{figure}
\centering \includegraphics[width=8.7cm]{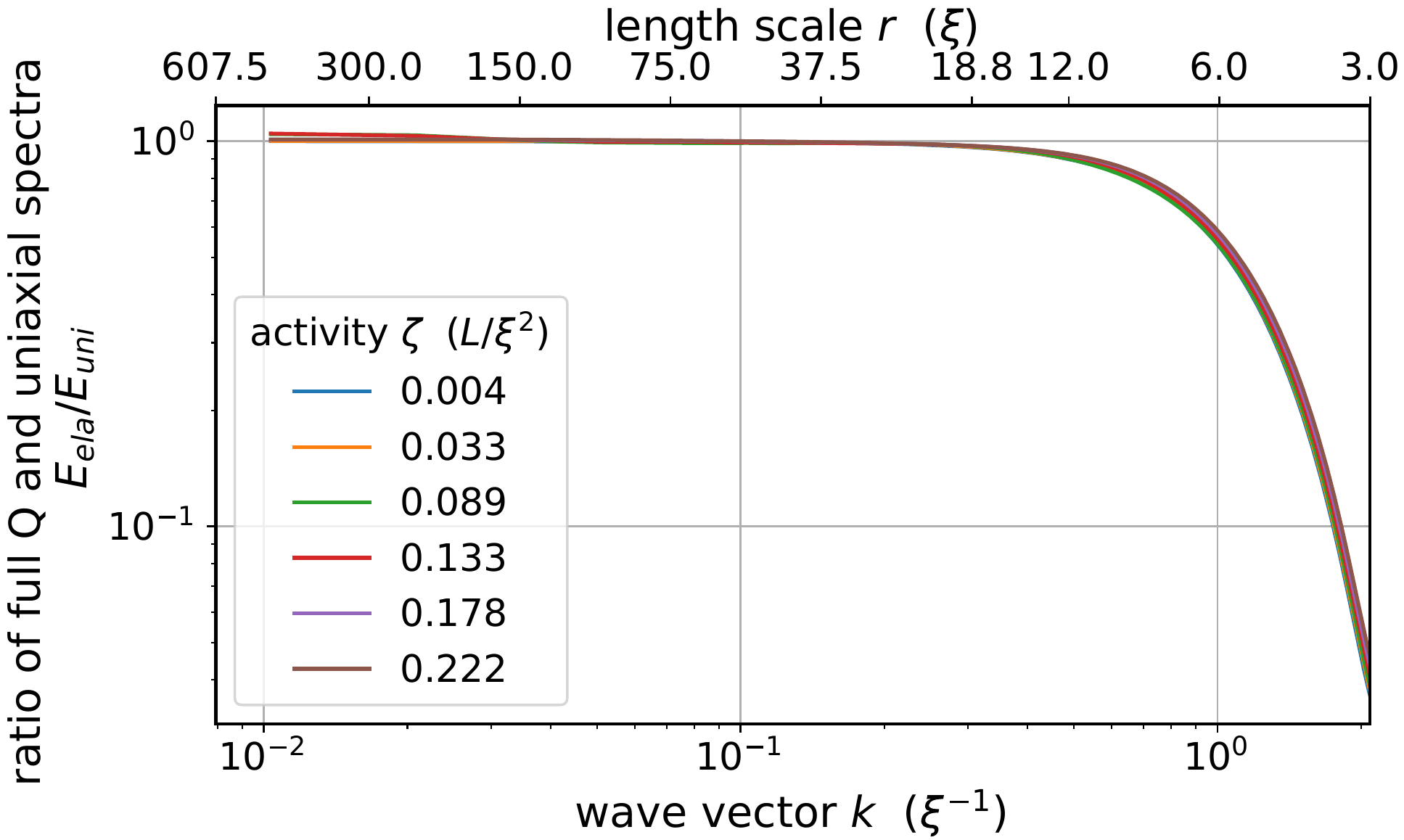}
\caption{Ratios of elastic energy spectra $E_{full}/E_{uni}$ obtained from using full tensor profiles $Q_{ij}$ and projected-out uniaxial tensor profile $Q_{ij}^{uni}$.}
\label{Fig_biaxiality}
\end{figure}

\begin{figure*}
\centering \includegraphics[width=\linewidth]{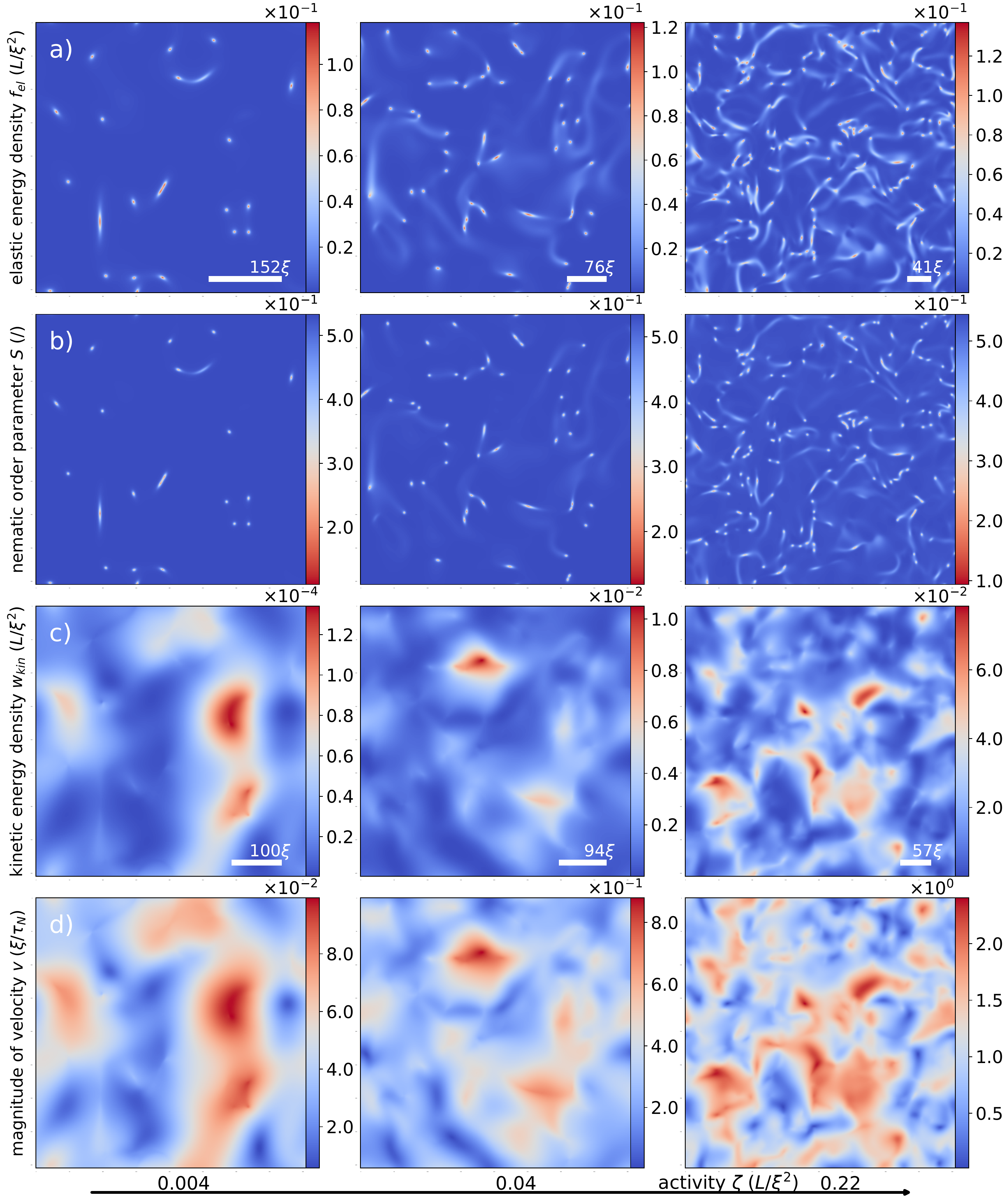}
\caption{Active chaotic (turbulent) patterns across different length scales for different activities. (a) Spatial structure of the nematic elastic energy density (given by $w_{ela} = \frac{L}{2} \partial_lQ_{ij} \partial_l Q_{ij}$), and (b) the corresponding profile of the nematic degree of order $S$. (c) Spatial structure of the kinetic energy density $w_{kin}$, and (d) the corresponding profile of the magnitude of the velocity field. Note the indicated lengths, as corresponding to the maxima in the elastic and kinetic energy density spectra. 
}
\label{Fig_extra}
\end{figure*}

\subsection{Energy spectra of 3D bulk active nematic turbulence}

Time averaged kinetic and elastic energy volume density of a dynamic steady state active turbulent state of three-dimensional active nematic, across the spectrum, and at different activities, are shown in Figure \ref{Fig2}. Kinetic energy density $E_{kin}$ (Fig.~\ref{Fig2}a) exhibits a maximum at scales of the order of hundreds of nematic correlation lengths, showing that the kinetic energy is largest at scales much larger than the typical size (thickness) of the topological defect lines. When approaching smaller length scales (i.e. larger $k$), the kinetic energy spectrum starts to decay  with increasing $k$ and shows an approximately $k^{-5}$ power-law dependence. Over the rough range of length scales from cca. $5$--$50\,\xi$ the kinetic energy density decreases strongly, for multiple (cca. five) orders in magnitude. This range of strong approximately power-law type of decrease is also evident from the spectral density of the enstrophy (Fig.~\ref{Fig2}c). The observed strongly non-monotonous dependence of kinetic energy (and enstrophy) indicates that there is a velocity lengthscale dictating the distribution of structures in the flow field. Varying the activity (from $0.004$ to $0.27\,L/\xi^2$) increases the kinetic energy density spectrum, as well as enstrophy density spectrum, across all length scales due to a broad increase of velocity magnitude. The increase in the kinetic energy density and enstrophy density is observed to scale  roughly as $\propto \zeta^2$, in a notable range of length scales, as shown in inset to Fig.~\ref{Fig2}c. Figure~\ref{Fig2}b shows the nematic elastic energy density spectrum $E_{el}$, which also exhibits a distinct maximum at all studied activities. Peak position  is activity dependent and shifts towards shorter length scales with increasing activity, inline with the anticipated  scaling $k_{peak} \propto \zeta^{0.5}$ (see Section~\ref{sec:active_lengthscale}).

The profile of the active nematic elastic spectrum at small scales (large wave vectors) is roughly independent of activity up to approximately $6$ nematic correlation lengths, which can be attributed to the effective thickness of the defect lines (i.e. defect cores), which is same at all activities. Increasing the activity effectively only multiplies this small scale (large-$k$) part of the spectrum, which can be understood by more defect lines being formed with increasing activity but of roughly same effective thickness. To provide further insight into this part of the elastic energy spectrum as belonging to molten defect cores, we compare the elastic spectra computed using the full $Q_{ij}$ tensor profiles (obtained directly from numerical calculations of the active turbulence) and the profile of uniaxial tensor $Q_{ij}^{uni}$ of the same data, calculated from the nematic degree of order $S$ and director $n_i$ as $ Q_{ij}^{uni} = \frac{3S}{2}(n_in_j - \frac{1}{3}\delta_{ij})$. Figure~\ref{Fig_biaxiality} shows that the full $Q_{ij}$ elastic spectra and uniaxial $Q_{ij}^{uni}$ are different at length scales below cca $10-20\xi$, notably independently of the activity. The difference between the spectra is the result of biaxiallity in the nematic profiles, which is non-negligible only in the defect cores. This analysis further supports that defect lines are of constant width, as explained before.

\subsection{Active nematic elastic lengthscale analysis}
\label{sec:active_lengthscale}
Figure~\ref{Fig_extra} shows---complementary to spectral analysis---the real space profiles of the kinetic and elastic energy densities of the active turbulence. The elastic energy density is highly localised (around the active defect cores) as shown in Fig.~\ref{Fig_extra}a and b, whereas the kinetic energy is more spread-out (Fig.~\ref{Fig_extra}c and d), which indeed were the general features seen in the Fourier spectra, but in reciprocal space with no direct real-space interpretation. In real space, the maxima of the elastic energy spectra are clearly related to location of the nematic defect lines, directly indicating that the position of the peak in the active nematic elastic energy spectra comes from the effective average separation of topological defects.  Complementary, as seen from Fig.~\ref{Fig_extra}, the maxima in the kinetic energy spectra roughly correspond to the typical separation between the effective local maxima (peaks) in the flow field.

\begin{figure}[h]
\centering \includegraphics[width=8.7cm]{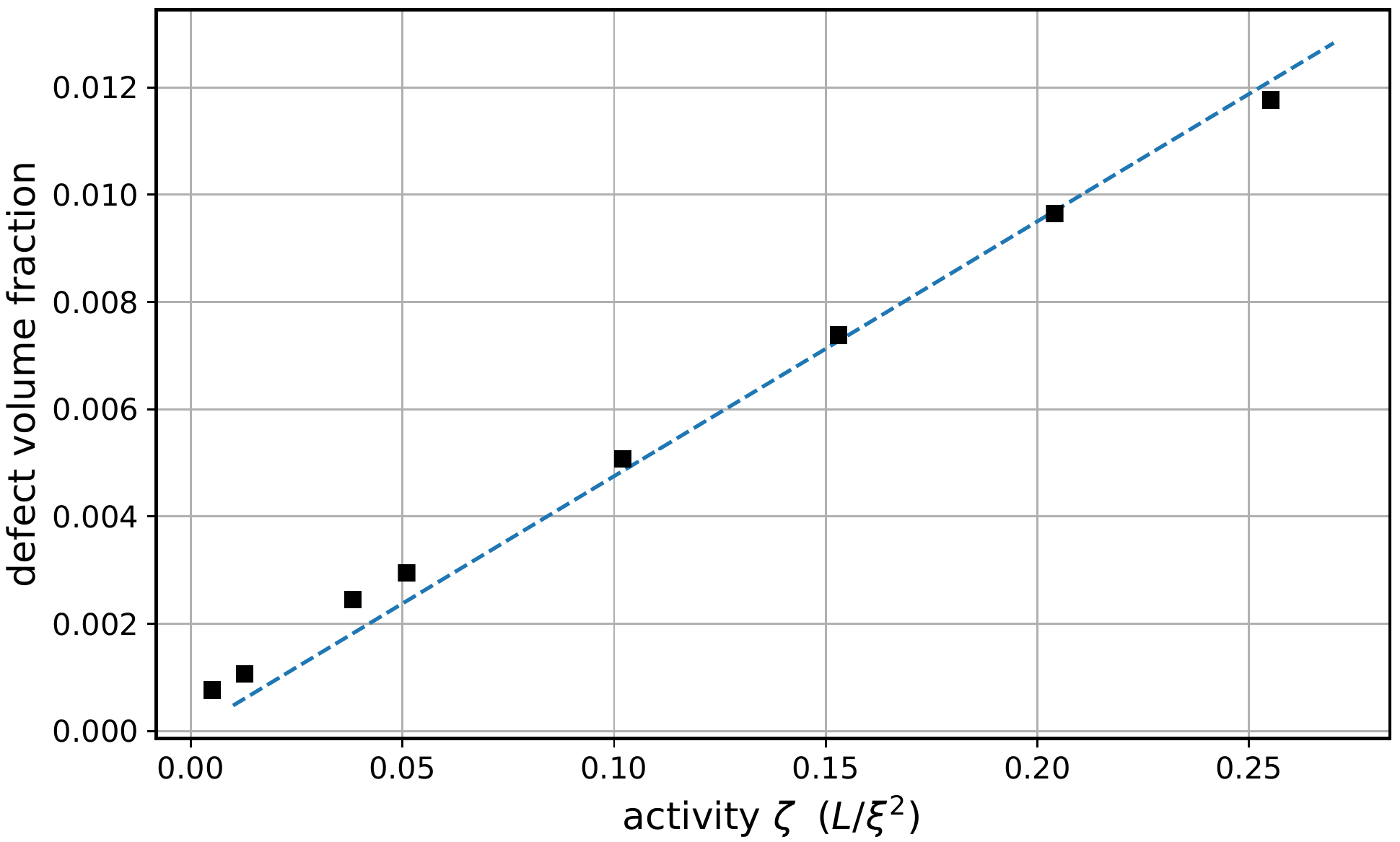}
\caption{Topological defect volume fraction as function of activity for dynamic steady state of active turbulence in three-dimensional active nematic. Results of numerical calculations and analysis are shown with black squares; dotted blue line shows linear curve $\propto \zeta$. The defect region is defined as volume where the nematic degree of order $S < 0.4$ (bulk equilibrium value $S_\text{bulk} = 0.533$).
}
\label{Fig3}
\end{figure}

The correspondence between the  typical lengths  obtained in active nematic elastic spectral analysis (i.e. the spectral maxima) and the effective average separation between the defect lines is independently tested by 
direct geometrical tracking of the whole defect line network. To geometrically determine the effective separation between the defects lines, the volume fraction of defect cores in a dynamic steady-state of active nematic turbulence is calculated, as shown for different activities in Fig.~\ref{Fig3}. Specifically, we identify defect cores as regions with reduced  nematic degree of order $S<S^\text{thr}$, where $S^\text{thr}= 0.4$ is used for results presented in Fig.~\ref{Fig3}. As a function of activity, the defect volume fraction in the dynamic steady state of the active turbulence is observed to increase approximately  linearly.

\begin{figure}[!hb]
\centering \includegraphics[width=8.7cm]{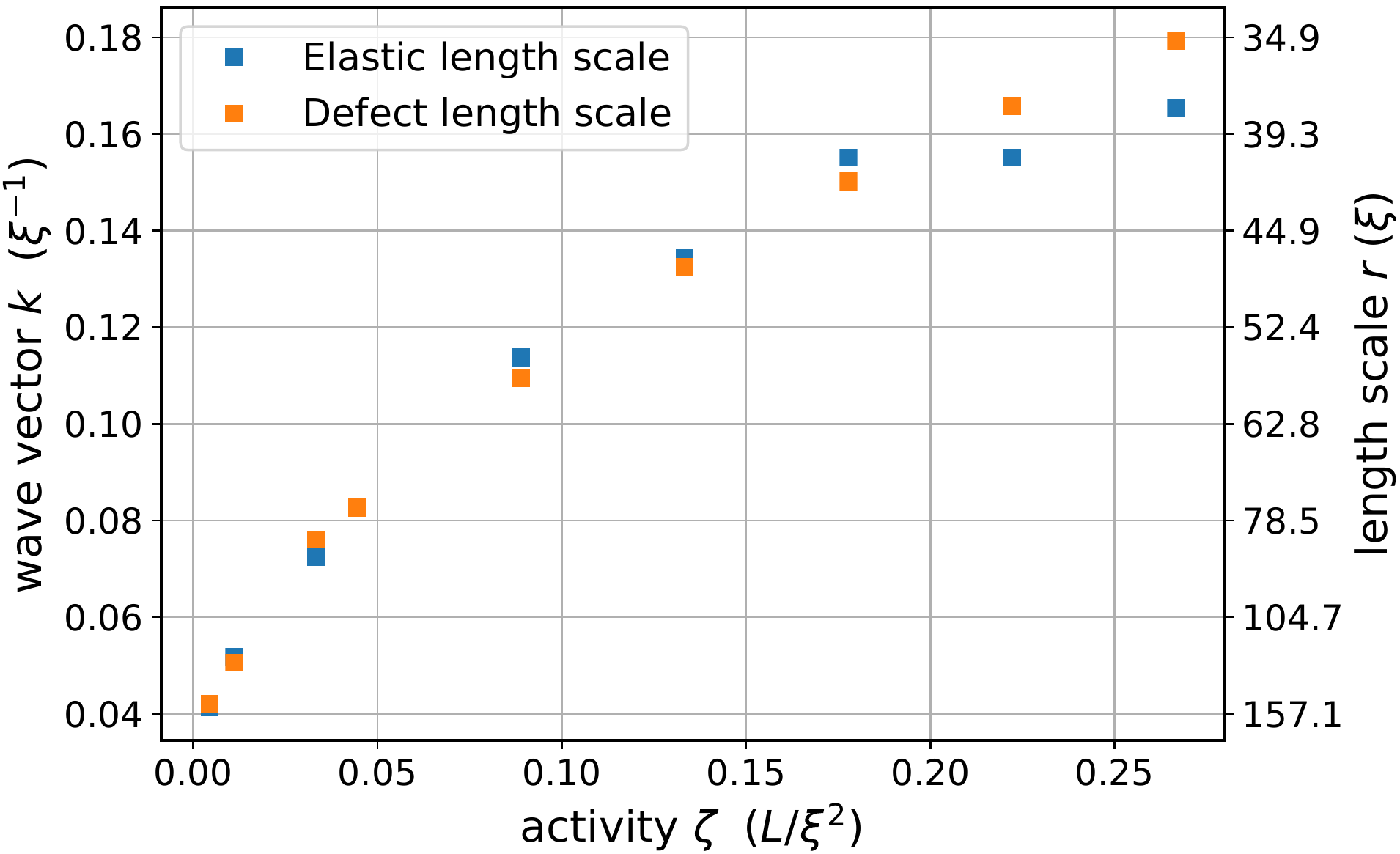}
\caption{Comparison of elastic length scale (blue) and defect length scale (orange). The constant prefactor of the defect length scale dependence is sensitive to changes of the used defect line cross section. A value of $d_{cs} = 4.1 \xi$ was used for the defect cross section diameter, based on the measured average.}
\label{Fig_comp}
\end{figure}

Defect volume fraction is proportional to the total length of the active nematic defect lines $x_d$, which in turn can be used to extract the effective length scale attributed to the defect-to-defect separation $l_d=\sqrt{V/x_d}$, where $V$ is the volume of the active nematic. Within the literature, the scaling of $l_d\propto \zeta^{-0.5}$ is predicted~\cite{HemingwayEJ_SoftMatter12_2016}. Such scaling explains a linear dependence of defect volume fraction on activity, which is appproximately observed in Fig.~\ref{Fig3}. Finally, the defect lengthscale $l_d$ obtained directly from the defect volume fraction calculation agrees very well with the values extracted from the elastic energy spectra, as shown in Fig.~\ref{Fig_comp}, indicating that indeed, the effective defect-defect separation performs as an effective length scale in the elastic behaviour of active nematic turbulence.

\subsection{Active nematic velocity lengthscale analysis}
The flow fields of the 3D active nematic turbulence can be analysed in the direct space --- complementing the mode analysis in Fig.~\ref{Fig2} --- by considering the autocorrelation functions of the velocity field, inline with the analysis of the 2D active nematic turbulence~\cite{GiomiL_PhysRevX5_2015}. We use the normalized, one-dimensional, zero-mean velocity autocorrelation function $C_{\mathbf{u}, \mathbf{u}}$ as:
\begin{align}
C_{\mathbf{u}, \mathbf{u}}(R) &= \frac{\oiint_{|\mathbf{R}| = R} \Big( \int \tilde{u}_i(\mathbf{r})\tilde{u}_i(\mathbf{r} + \mathbf{R})d^3 r \Big) dS_{\mathbf{R}}}{\int \tilde{u}_i(\mathbf{r}) \tilde{u}_i(\mathbf{r})d^3 r} ,\\
\tilde{u}_i(\mathbf{r}) &= u_i(\mathbf{r}) - \frac{1}{V}\int u_i(\mathbf{r}) d^3r,
\label{utilde}
\end{align}
where bulk integration is performed over the whole system volume $V$ and the surface integral recomputes the full three-dimensional correlation into a one-dimensional radial correlation by averaging over a thin spherical shell, similarly as in Eq.~\ref{discrete}.  
Note that subtracting the mean in Eq. (\ref{utilde}) only changes the correlation function by a constant factor, where the mean of the velocity field is typically two orders of magnitude smaller then its standard deviation at all activities, $\langle u \rangle / \sqrt{\langle u^2 \rangle} \ll 1$.
The normalization and zero mean ensure that the values of the autocorrelation function will be confined to the interval $[-1, 1]$. The velocity autocorrelation is computed via the Fourier convolution theorem.

\begin{figure}[h]
\centering \includegraphics[width=8.7cm]{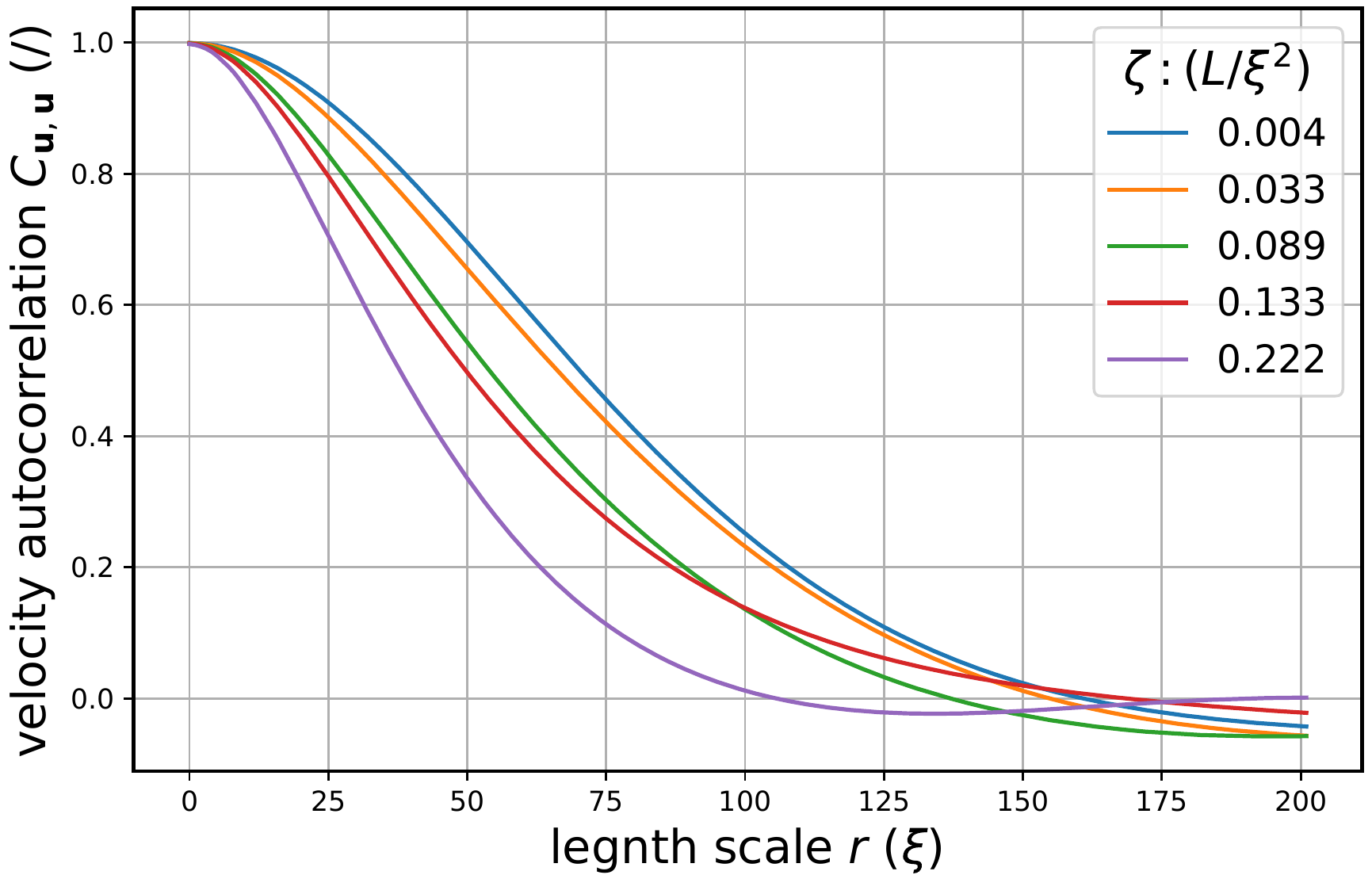}
\caption{Velocity autocorrelation function  of active turbulence in 3D active nematics for various activities.}
\label{Fig_corrf}
\end{figure}

The velocity autocorrelations of the flow fields of 3D active nematic turbulence are shown for different activities in Fig.~\ref{Fig_corrf}. The general trend is that the range of the correlation decreases with increasing activity. The calculated correlation function could be used to provide insight into characteristic lengths in the velocity field of the 3D active turbulence.  Selection of the actual length can be me made by selecting some distinct cut-off value of the correlation function, possibly, also in combination with the peak-values of the kinetic energy or enstrophy spectra.

\subsection{Active turbulence in external aligning field}

\begin{figure}[!h]
\centering \includegraphics[width=7.8cm]{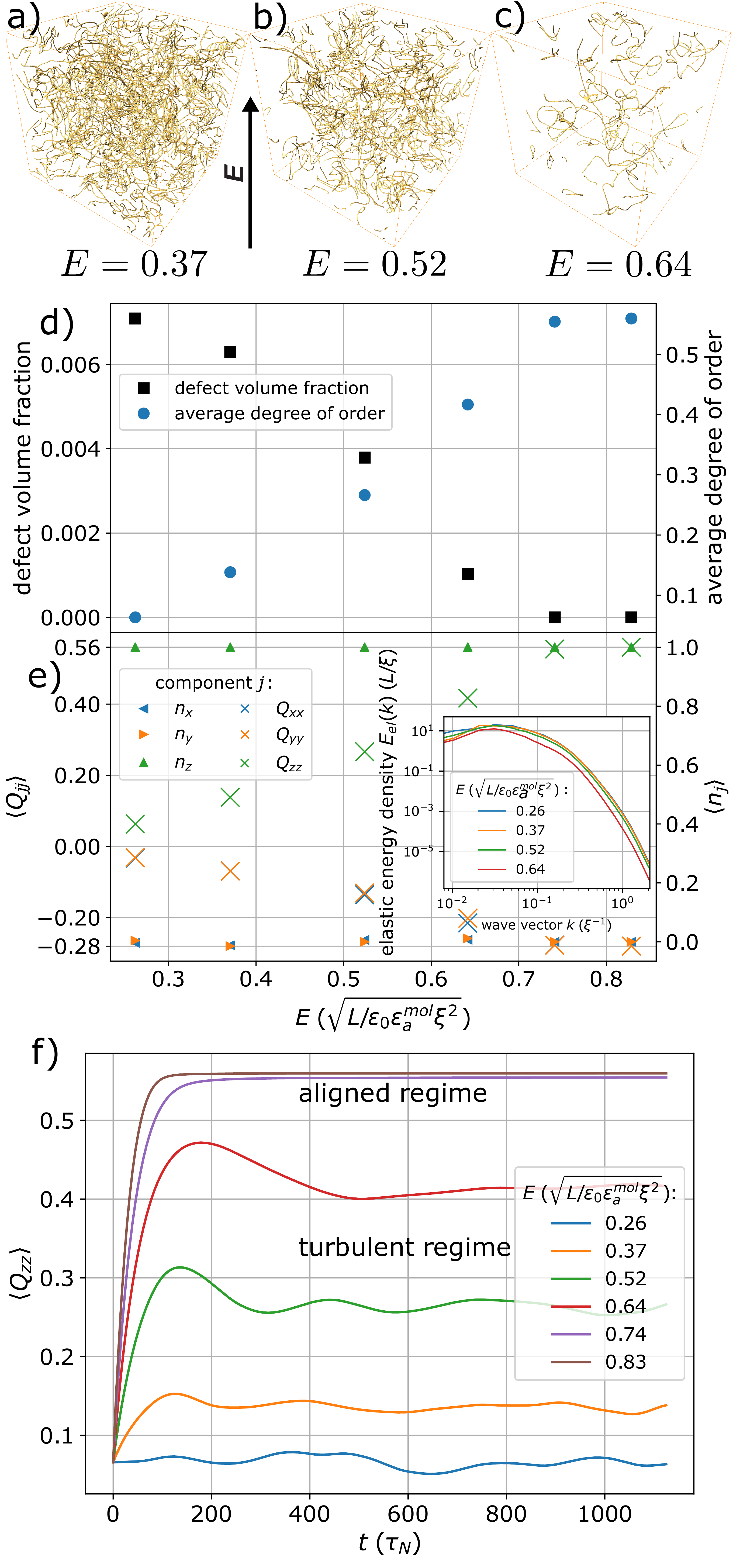}
\caption{Effect of external homogeneous aligning (electric) field on 3D active nematic turbulence at activity $\zeta = 0.13 L/\xi^2$. 
(a,b,c) Structure of 3D topological defects for increasing field strengths, $E=$ $0.37$, $0.52$, $0.64$  (in units of $\sqrt{L/\epsilon_0 \epsilon_a^{mol} \xi^2}$). Defect lines are drawn in yellow as isosurfaces of the nematic degree of order at $S = 0.35$. (d) Variation of defect volume fraction and average degree of order on the field strength. (e) Variation of the average principal Q-tensor components and components of the director on the field strength $E$. Inset shows elastic spectra essentially unaffected by electric field for $E < 0.7$ (i.e. below transition). (f) Time-dependence of average $<Q_{zz}>$ for different field strengths $E$, upon switching on the external field, starting from 3D active turbulence dynamic steady state.}
\label{Fig_electric}
\end{figure}

We have investigated the effect of an aligning homogeneous (electric) field on an the 3D active nematic turbulence, with results shown in Figure~\ref{Fig_electric}. In the considered geometry and coupling regime (see Eq.~\ref{free_en}; we assume positive dielectic anisotropy), the active nematic director is energetically favoured to align along the direction of the external field ($z$ direction) with an external torque arising from Eq.~\ref{free_en}; note that such aligning effect could possibly be achieved also with other fields, like magnetic, mechanical, or light fields, which is well known in passive nematic complex fluids~\cite{book-deGennes}. Note also, that for the presented geometry of active nematic (box with 3D periodic boundary conditions), there is no preferred direction for the active nematic director or pinning of the active nematic to a surface; therefore, the \textit{average} director (in time or in space) of the 3D active turbulence is wholly aligned along the field direction even for (in principle) arbitrarily small field strengths. 

The density of defects in the 3D active nematic turbulence decreases with increasing field strength as shown in Fig.~\ref{Fig_electric}a--c, which in turn results in an increase of the average degree of order (Fig.~\ref{Fig_electric}d). When the electric field strength crosses a critical activity dependent value $E_c \approx 0.7 \sqrt{L/\epsilon_0 \epsilon_a^{mol} \xi^2}$ (for $\zeta = 0.13 L /\xi^2$), we observe a transition of the characteristic 3D active turbulence defect pattern to a completely field-aligned configuration of the nematic, with no topological defects and homogeneous nematic director along the field direction. This transition behaviour is clearly reflected in the plateauing of the $Q$ tensor and derived quantities (Fig.~\ref{Fig_electric}e). Remarkably, the elastic spectrum (inset of Fig.~{\ref{Fig_electric}e}) is similar for a range of fields, for $E < E_c$, but drops abruptly to effectively zero above the transition $E > E_c$. 

The actual application (i.e. switching on) of the aligning field causes time dependent transient realigning and structural changing in the 3D active turbulence, as shown in Fig.~\ref{Fig_electric}f. Upon switching  on the field (from zero to constant value), effective oscillations in the average ordering of the 3D active turbulence, shown by $\langle Q_{ii}\rangle$, are observed for field strengths below the threshold ($E < E_c$) and monotonous approach to the steady state is seen for $E > E_c$. The electric fields of moderate strength $E \lessapprox E_c$ effectively over-drive the nematic, which is then dissipated to a dynamic steady-state. An analogous process in electric field driven passive nematic fluids is known as the optical bounce and is of major importance in display design \cite{vanDoornCZ_JApplPhys46_1975}. 

\section{Discussion and conclusions}
We have performed the spectral energy analysis of 3D bulk active nematic turbulence where activity is manifested in a dense tangle of spatially and time-varying active defect lines. We calculate the elastic energy spectrum of active nematic order parameter and kinetic energy spectrum of the flow field, showing that the active nematic elastic energy is concentrated on scales comparable to the average distance between defect lines, typically of 10-100 nematic correlations lengths dependent on activity, whereas the kinetic energy is largest at somewhat larger scales of  typically several 100 nematic correlation lengths. We show that the effective defect-defect separation scales with activity as $\sim\zeta^{0.5}$. The role of nematic biaxiality is addressed showing that it is roughly activity independent in the active turbulence but becomes energetically important at scales comparable with defect core size. Velocity autocorrelation is also discussed in view of interpreting the kinetic energy spectra. We also explore how the 3D active nematic turbulence responds to a homogeneous external aligning field and show that it aligns the average active neamtic director field  along the aligning field axis, despite the active nematic remaining in the active turbulent regime. As the strength of the electric field is increased, the degree of alignment also increases, and above a certain activity-dependent electric field threshold, full alignment with homogeneous director is observed. We also observe that the transition dynamics -as the aligning field is turned on- can possibly lead to transient (damped) dynamic oscillations in the active turbulence. 
 
A span of diverse active materials---from active biological fibres to bacteria---show the regime of active turbulence, but it is an open interesting question if or to what extend are these seemingly similar irregular material dynamic regimes really equivalent. Using spectral analysis of the main energetic contributions in the system, as shown in this study, could possibly provide such basis for comparison, comparing the characteristic scalings of the energy terms (e.g. with the activity), in view, as if searching for the \quotes{universality classes} of the active matter. As only one consequence, establishing such more unifying understanding of diverse active materials would allow active matter science to move even further from system-specific models and formulate  general concepts and approaches, also maturing the field, and making it further accessible to technological ideas and applications. 
More generally, spectral energy analysis could provide an interesting robust method for mapping experiments to different models of active nematics. One might expect models to separate into classes, characterized by differing scaling laws for various physically motivated quantities, unperturbed by variations of model parameters away from critical values -- as also just recently discussed in~\cite{alert2019universal,bourgoin2019kolmogorovian}. Experimental spectra could then be matched with appropriate theoretical models in a parameter-free manner.

\section{Acknowledgements}
Authors acknowledge funding from Slovenian Research Agency ARRS grants P1-0099, N1-0124, J1-1697 and  L1-8135. M.R. also acknowledges support under EPSRC Grant No. EP/R014604/1 at Isaac Newton Institute, University of Cambridge (\quotes{The mathematical design of new materials} program).

\section{Conflicts of interest}
There are no conflicts to declare.

\providecommand*{\mcitethebibliography}{\thebibliography}
\csname @ifundefined\endcsname{endmcitethebibliography}
{\let\endmcitethebibliography\endthebibliography}{}


\end{document}